\documentclass[conference]{IEEEtran}  
\usepackage{graphicx}
\usepackage{caption}
\usepackage{multirow}
\usepackage{morefloats}
\title{
A Big Data Analysis Framework Using \\Apache Spark and Deep Learning
}

\author{ \parbox{2.5 in}{\centering Anand Gupta \\
		Dept. of Computer Engineering\\
        NSIT, University of Delhi\\
        Delhi, India\\
        {\tt omaranand@nsitonline.in}}
        \parbox{2.5 in}{\centering Hardeo Kumar Thakur \\
        Dept. of Computer Science\\
        Manav Rachna University\\
        Delhi, India
        {\tt hkthakur@mru.edu.in}}
        \\\\\\
        \parbox{2.35 in}{\centering Ritvik Shrivastava \\
        Dept. of Information Technology\\
        NSIT, University of Delhi\\
        Delhi, India\\
        {\tt ritviks.it@nsit.net.in}}
        \parbox{2.35 in}{\centering Pulkit Kumar \\
        Dept. of Information Technology\\
        NSIT, University of Delhi\\
        Delhi, India\\
        {\tt pulkitk.it@nsit.net.in}}
        \parbox{2.35 in}{\centering Sreyashi Nag \\
        School of Computer Science\\
        Carnegie Mellon University\\
        Pittsburgh, PA, USA\\
        {\tt sreyashn@andrew.cmu.edu}}
}

\begin{document}

\maketitle
\thispagestyle{empty}
\pagestyle{empty}

\begin{abstract}

With the spreading prevalence of Big Data, many advances have recently been made in this field. Frameworks such as Apache Hadoop and Apache Spark have gained a lot of traction over the past decades and have become massively popular, especially in industries. It is becoming increasingly evident that effective big data analysis is key to solving artificial intelligence problems. Thus, a multi-algorithm library was implemented in the Spark framework, called MLlib. While this library supports multiple machine learning algorithms, there is still scope to use the Spark setup efficiently for highly time-intensive and computationally expensive procedures like deep learning. In this paper, we propose a novel framework that combines the distributive computational abilities of Apache Spark and the advanced  machine learning architecture of a deep multi-layer perceptron (MLP), using the popular concept of Cascade Learning. We conduct empirical analysis of our framework on two real world datasets. The results are encouraging and corroborate our proposed framework, in turn proving that it is an improvement over traditional big data analysis methods that use either Spark or Deep learning as individual elements.
\end{abstract}
\vspace{4mm}

\section{INTRODUCTION}

\subsection{Overview}
With the amount of data growing at an exponential rate, it is necessary to develop tools that are able to harness that data and extract value from it. Every organization or company, be it in the healthcare, manufacturing, automobile or software industry, needs to manage and analyze the large data volumes that it generates. This in turn leads to faster and more efficient operations. 
The growing need to manage increasingly massive data collections has led to significant interest in developing such big data tools. Research is being conducted in the areas of Big Data Infrastructure (cloud computing, energy-efficient computing, software systems, new programming modules etc), Big Data Management (search and mining, algorithmic systems, data acquisition, integration and cleaning, computational modeling, graph mining, distributed and peer-to-peer search etc), Big Data Search and Mining (social network analysis and mining, web search, semantic based data mining etc), Big Data Security (high performance cryptography, privacy threats to big data, sociological aspects of big data privacy etc) and many other related fields. This steadily growing demand has led to the development of several big data analytics tools in the industry. 
 
The need for big data analytics frameworks can be seen when implementing any algorithm on large datasets. Any such process uses a single core of the CPU in a local system. To improve performance as the data size increases, GPUs are being increasingly used which contain multiple cores. Thus, parallel processing can be easily done due to their distributed architecture. But because GPUs are not always economically feasible and accessible, there is a need for a framework that uses the existing CPU in the local system in a distributed environment. 
 
Among the most prominent tools that achieve this task is Hadoop, an open source platform that provides excellent data management provisions. Its primary use is to facilitate the processing of very large datasets in a distributed computing environment using components like the Hadoop Distributed File System (HDFS), MapReduce, HBase, Hive etc. This paper however, explores a more efficient and robust tool, Apache Spark, that was designed to work in conjunction with Hadoop in order to address some of its limitations. 
 
\subsection{Challenges}
The primary purpose of Spark is to efficiently handle data of a large magnitude. Using built-in libraries individually, existing systems have managed to use Spark for such analysis of Big Data. However, there is still scope for exploring new models that yield results with greater accuracies while still using the Spark framework to remain computationally feasible. The key questions are:
(1) How to extract maximum information from the pre-existing features in the dataset to yield optimum accuracy?
(2) How to effectively address the problem of class imbalance that is present in large scale real world datasets?
(3) How to incorporate the recent advances made in the field of Artificial Intelligence while still utilizing Spark$'$s computational power?
 
\subsection{Countering the Challenges}
To address the challenges mentioned above, this paper presents a framework that harnesses the power of cascade learning under the umbrella of Spark$'$s framework. Using this technique, the knowledge gained from the primary architecture is used to add value to the pre-existing features of the data being analyzed. This improved information is now fed into a second learning architecture. The output of this two-step learning process is a model that is significantly more accurate than traditional single stage machine learning models. Further, the resultant model also addresses the class imbalance problem. 
 
\textit{Problem Definition:} Given (any) large scale real world dataset, the task is to propose a framework which can give a highly accurate and efficient prediction model.

\subsection{Applications}
Applications of the proposed framework lie in multifarious domains. A few of these have been discussed next:
(1)	Healthcare - Analysis of detailed large-scale equipment readings in the form of both texts and images. Good learning algorithms will result in better healthcare predictions and recommendations. We study one such dataset in this paper.
(2) Education - Analysis the performance of students, teachers and supporting staff.
(3) Physical Capital based Industrial Sector - Manufacturing and Instrumentation details such as data recordings for large MNCs can be studied using this framework.
(4) IT Industry - Improving existing systems by replacing or appending manual work with big data based artificial intelligence processes. 

\subsection{Structure of the Paper}
The overall framework presented in this paper aims to tackle the aforementioned challenges. The paper is divided as follows:
Section II is a review of the literature that is available in the areas under consideration and recent advances in the same.
Section III provides a detailed explanation of the proposed approach and the reasoning behind it.
Section IV is a detailed study of the experiments, simulations and results that are obtained during the course of this paper’ preparation.
Sections V and VI provide conclusions to our research along with the scope and possibilities for future work.

\section{RELATED WORK}The literature in this area includes work that has been done in the fields of big data, Spark, machine learning, deep learning and cascade learning. We study this literature with the intention to analyse past work in these areas and understand their limitations in order to engineer a system capable of handling these limitations.
The literature survey pertaining to this paper has been presented in four parts: (i) Work related to Apache Spark, (ii) Work related to feature set and data manipulation, (iii) Work related to deep learning using multi-layer perceptrons and (iv) Work related to cascade learning.

\subsection{Work Related to Apache Spark}
The architecture and utility of Apache Spark was first presented to the community by the authors of \cite{zaharia2010spark}. It gives a brief overview regarding the programming model, which includes RDDs, parallel computing etc. It also introduces a few implementations in the environment. \cite{meng2016mllib} introduces Apache Spark’s machine learning library, MLlib. This includes its core features as well as the different components constituting the MLlib library. The work done in \cite{fu2016spark} analyzes Spark’s primary framework by running a sample ML instance on it. The authors present comprehensive results based on their evaluations that highlight Spark’s advantages. The authors of \cite{nair2015streaming} present a similar usage of Spark by analyzing twitter data using Spark’s framework. The massive size of the twitter dataset used highlights the significance of using a tool like Spark. The authors of \cite{nodarakis2016large} have performed similar twitter sentiment analysis using Apache Spark. 

\subsection{Work Related to Feature Set and Data Manipulation}
The class imbalance problem is a significant area of interest in the machine learning community that we have addressed in this paper. The authors of \cite{kotsiantis2006handling} present a review of various tools and techniques that can be used to solve the imbalance issue. A comprehensive analysis of various methods targeted towards solving the problem has been mentioned in \cite{sonak2016new}. A major part of the analysis presented in this paper involves the manipulation and extraction of relevant features from both the datasets. This kind of feature engineering can be seen in \cite{guyon2003introduction}. The authors of \cite{popescu2014feature} study feature extraction and feature selection for machine learning in image classification. \cite{dey2016paraphrase} presents a state-of-the-art solution to the natural language processing problem of paraphrase detection, using explicit feature set construction. Authors of \cite{lavravc2010explicit} also present a comprehensive example illustrating explicit feature construction and manipulation for covering rule learning algorithms.

\subsection{Work Related to Deep Learning using MLPs} Deep learning using multiple layer perceptrons has been fast gaining significance in the artificial intelligence community, especially when handling massive data. The authors of \cite{silva2008data} introduce data classification using multi-layer perceptrons. \cite{sharma2014big} presents an instance of large scale analysis conducted on big data using MLPs. This highlights the advantages offered by MLPs when dealing with large datasets instead of small or moderately-sized ones. Another instance of the usage of MLPs in classification problems is presented in \cite{hu2010pattern}. This method used a fuzzy integral based activation function. Backpropagation is primary method of learning used in multi-layer perceptrons.  The authors in \cite{pal1992multilayer} present a fuzzy neural network model based on MLP using the backpropagation algorithm which illustrates how learning takes place in such a multi-layer neural network. 

\subsection{Work Related to Cascade Learning}
The two stages of the framework mentioned in this paper, namely analysis using Spark and multi-layer perceptron, is connected via cascading. Cascading or Cascade Learning is typically used for cases where the classes are heavily imbalanced and a suitable inference cannot be gathered from the data. This approach has been previosuly applied in diverse fields such as Natural Language Processing, Computer Vision etc. and has proven to be quite effective. \cite{sheikh2015} proposed a time efficient two-stage cascaded classifier for the prediction of buy sessions and purchased items within such sessions. \cite{simon2016} proposed a cascade of feature-sharing deep classifiers, called OnionNet, where subsequent stages added both new layers as well as new feature channels to the previous ones. Recently, \cite{christ2016automatic} cascaded two fully convolutional neural networks \cite{long2015fully} for a combined image segmentation of the liver and its lesion and produced state-of-the-art results on the task.

Having studied the existing literature, we aim to address the aforementioned challenges through the framework described in this paper. In the following section, we present a detailed study of the proposed framework and the techniques incorporated in it.
\\

\section{PROPOSED APPROACH}
In this section, we describe in detail the framework mentioned in this paper. Our approach combines the benefits of using a big data processing framework like Spark along with the advantages of deep learning on large datasets by using an approach called Cascade Learning. This technique is discussed below followed by the structure of the framework used in this paper.

\subsection{Cascading}
In traditional supervised machine learning algorithms, individual models are trained for their respective tasks. This has been a highly successful technique whenever the labeled data for the task at hand is available. Although, in situations where there is a lack of adequate labeled data corresponding to the use case, the traditional algorithms result in unreliable models. \\

Cascading allows for the application of `knowledge' gained from a previously trained model to another model, so as to improve the performance on the task at hand. This model-to-knowledge transfer is used as an additional feature for the overall model, hence providing extra relevant information. \\\\
\textbf{Definition:} \textit{Cascade learning} or \textit{cascading classifiers} can be defined as a subtype of ensemble learning, where the output of one classifier is fed into another classifier as input. This in turn enhances the overall classification performance.

\subsection{Framework}
The framework described in this paper is centered around solving real-world big data problems using big data analytics frameworks and artificial intelligence. To tackle such problems is challenging due to time and space constraints. Due to the sheer enormity of data and the lack of machines with high computation power, there is a need for supportive learning frameworks that can handle such data using just the stipulated resources. 

The model presented in this paper attempts to overcome these challenges. It combines the ideas of big data analysis, machine learning, deep learning and cascading. The core structural framework presented in this section forms the crux of our research and experiments. 

The essence of this framework lies in using cascading between the aforementioned processes of big data analysis and deep learning. These processes are explained next.\\

\subsubsection{Big Data Analysis Using Spark}

The various machine learning algorithms on Spark have been studied in depth with reference to big data. For the purpose of this research, we use the MLLib library of Spark to implement Logistic Regression, Decision Trees and Random Forest regression algorithms. In this stage, the pre-processed dataset is passed through these algorithms to create a regression model that presents the probability of each data-point belonging to a binary class. This stage is a binary-learning stage.\\

\subsubsection{Cascading}
As explained in Section III(A), Cascading involves using the knowledge obtained from one model to train another related model. In this framework, we engineer a modified version of the original dataset by appending the probabilities that were obtained through stage 1. This gives each data point an attribute that closely resembles the ground truth, in turn establishing a strong distinguishing feature in the dataset. This modified-dataset is our `knowledge', which is used as an input for stage 3.\\

\subsubsection{Deep Learning}
This stage is the culmination of our framework. The 	`knowledge' obtained in the form of the modified dataset from previous stages is used to train a multi-layer perceptron (MLP) architecture. The exact architecture for this layer is defined according to the application that it is being used for. This stage is suited for both binary and multi-class learning, according to the requirements of the application. A MLP takes into consideration the depth of the network according to the complexity of the problem and the system’s computational complexity. \\

\begin{figure*}
\includegraphics[width=\textwidth]{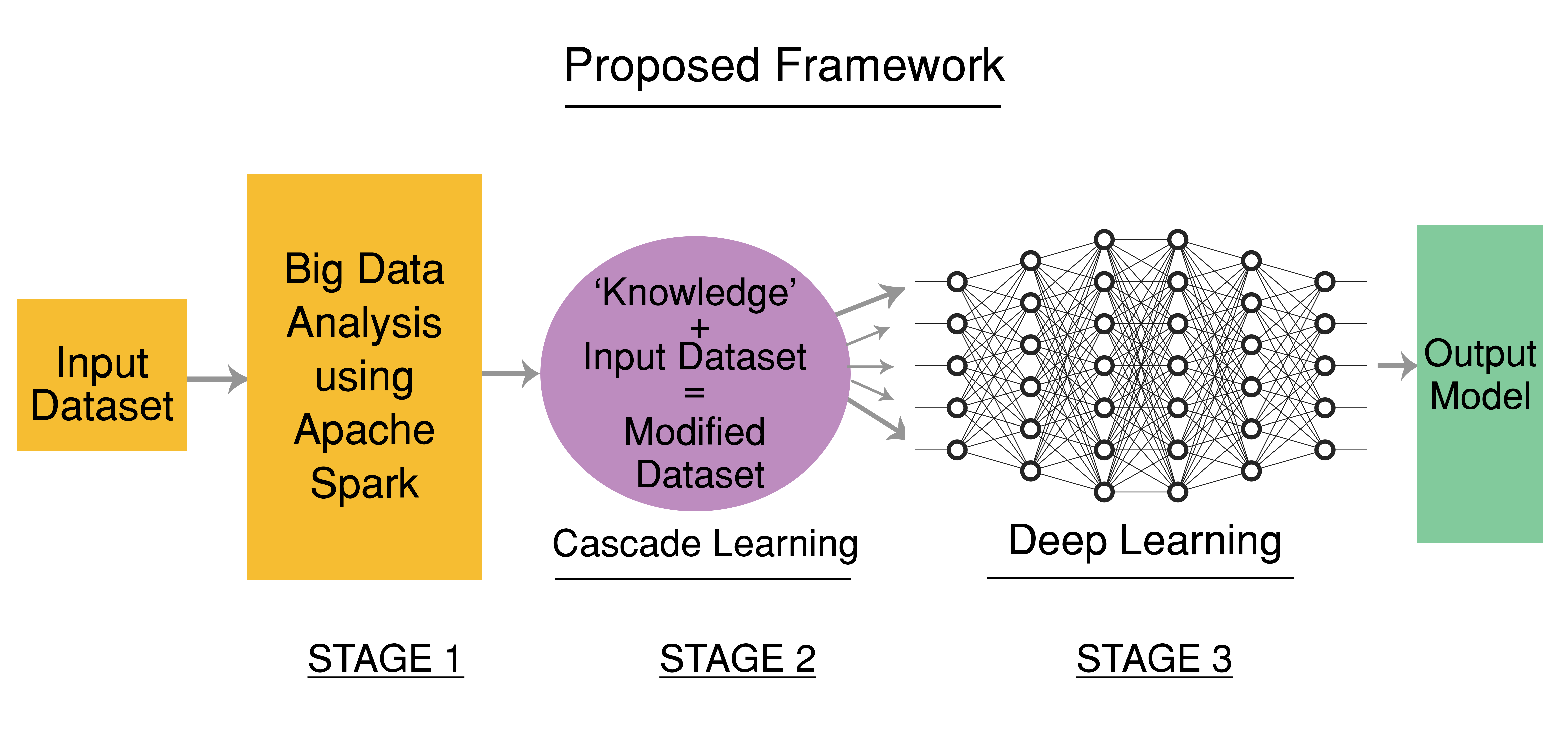}
\caption{Schematic Representation of the Proposed Framework}
\label{Proposed Framework}
\end {figure*}

The framework is detailed in Fig. \ref{Proposed Framework}.

\subsection{Steps of the Framework}
\textbf{Stage 1 : Big Data Analysis using ML in Spark}
\begin{enumerate}
\item Input – Pre-processed dataset in the form of a RDD
\item Convert RDD to DataFrame (DF)
\item Read Features and Labels from DF
\item One Hot Encoding of the non-numeric features
\item String Indexing of each encoded feature
\item Vector assembly of one-hot-encoded features and numeric features
\item Convert the assembled vector into a Pipeline
\item Fit and Transform the Pipeline into a suitable form for Spark to read
\item Train the model using MLLib based features using the training data
\item Test on the whole data to obtain a binary prediction value of the label (the prediction can be defined according to the needs of the user)
\end{enumerate}

\textbf{Stage 2 : Cascading}
\begin{enumerate}
\item Append the predictions data to the original dataset file.
\item This is the creation of `knowledge' that will be used.
\item Use the modified dataset (`knowledge') for training in stage 3.
\end{enumerate}

\textbf{Stage 3 : Deep Learning}
\begin{enumerate}
\item Train a multi-layer perceptron (MLP) using the `knowledge' obtained in Stage 2, step 3.
\item This MLP can be produced by either, repeating steps 2-8 in stage 1 and replacing the ML approach with an MLP that is created using the internally defined library of Spark, or the MLP can be generated by creating the Artificial Neural Network from scratch.
\item For the purpose of Deep Learning and high quality training, we create a back-propagation network to continuously train the network and reduce the error in prediction.
\end{enumerate}

Having presented the framework and its components in detail, we next discuss some of the applications that this model can be used for. \\

The abovementioned framework has applications in various domains of Big Data analysis and Machine Learning, such as classification systems and recommendation engines.

\subsection{Reasoning Behind Choice of Approach}
The framework we have discussed in the earlier section is a novel method for solving traditional machine learning problems. It considers all aspects of a traditional machine learning while improving accuracy and speed of the setup. The major reasons behind choosing this framework are further discussed in detail.

\subsubsection{Enhanced Feature Set}
The modified dataset that is obtained in stage 2 enhances the existing feature set presented in any problem. It boosts the accuracy of any model where this data is used henceforth.
For example, in the H-1B Visa Application dataset, we obtain the probabilities of acceptance or denial of the application for every application. We append this probability to the original dataset to obtain the modified dataset or `knowledge' which gives the dataset a new indispensable feature. This feature emulates to some extent the ground truth of this data, hence improving the chances of high accuracy when the deep learning model is created and applied on it. In this manner, our framework is able to enhance the existing featureset to a better version of itself.

\subsubsection{Computational Time Consideration}
Running a two-layer approach using two deep learning models simultaneously is a tedious and time-draining task. Replacing a stage with a much faster big-data analytics framework such as Spark or Hadoop reduces the time complexity. The state-of-the-art MLLib library in Spark does not reduce the quality of the model while maintaining much shorter computational times as compared to traditional approaches. 
For example, if running the deep learning model twice on an over 3 million data-point dataset such as H-1B Visa Application dataset takes 6 hours on a machine, a combination of our model will do it in 3-3.5 hours with noticeably higher efficiency. Our framework can reduce computation times up to half in cases, which is extremely valuable when talking about big data analysis and computation.

\subsubsection{Continuous Learning and Improvement}
Given the second stage of the model is a back-propagation based deep learning architecture, this framework will create and learn new features on its own without explicit hand-crafting of features. This enables the model to work as a more reliable model and move closer towards the original ideology behind Artificial Intelligence.
For example, in the H-1B Visa Application Dataset, the process of learning through back-propagation keeps training the model repeatedly to improve the accuracy of prediction or any other supervised or semi-supervised task. As more data keeps coming in, this process of continuous learning will allow the model to actually help the applicants by generating a fair prediction on their chances of being approved. 
Whereas, in the Arrhythmia dataset, the model can keep learning new features with continuously improving dataset as well as the self-improving deep learning structure.\\

Having described our proposed framework in detail along with the motivation for the choice of approach described in this paper, we conduct a series of experiments to validate our model. This is presented in Section IV.

\section{EXPERIMENTATION}

The overall aim of experimentation is to perform a qualitative and quantitative analysis on the performance of the framework proposed in this paper. We begin by describing the system specifications of all the distributed server systems that are used to implement Spark and the deep learning architecture: \\

\begin{enumerate}
\item System 1

\begin{itemize}
\item Ubuntu 14.04.5 LTS  $(GNU/Linux$ 3.19.0-25-generic x86-64$)$
\item Intel Core i7 CPU $@$ 3.40GHz
\item 16GB RAM
\end{itemize}

\item System 2
\begin{itemize}
\item macOS Sierra 10.12.4 
\item Intel Core i5 CPU $@$ 2.8 GHz 
\item 8 GB RAM \\\\
\end{itemize}
\end{enumerate} 

We conduct our experiments on two real-world datasets:
\begin{enumerate}
\item \textbf{H-1B Visa Applications Record Dataset}\footnote{kaggle.com/nsharan/h-1b-visa} This dataset is a collection of over 3 million H-1B Visa applications that were filed in the United States of America during the time period of 2011-2016. The information regarding the applications was collected from all fifty states in the United States of America in addition to three major federal provinces. This data can be considered as a comprehensive and exhaustive representation of the H-1B application record filed in the US during this time period. 

\item \textbf{Cardiac Arrhythmia Dataset} This dataset consists of data corresponding to the Electrocardiogram - ECG - measurements of 452 patients.
\end{enumerate}

To validate the efficacy of our framework we conduct experimentation on four tasks. Three tasks are performed on dataset 1 and one task on dataset 2. 

First, we discuss the tasks run on dataset 1 or the \textbf{H-1B Visa Application Record Dataset}: 

\subsection{Task 1 - Classification on the basis of ``Case- Status"}
\vspace*{-2mm}
\begin{figure}
\hspace{-1mm}
\includegraphics[width=0.5\textwidth]{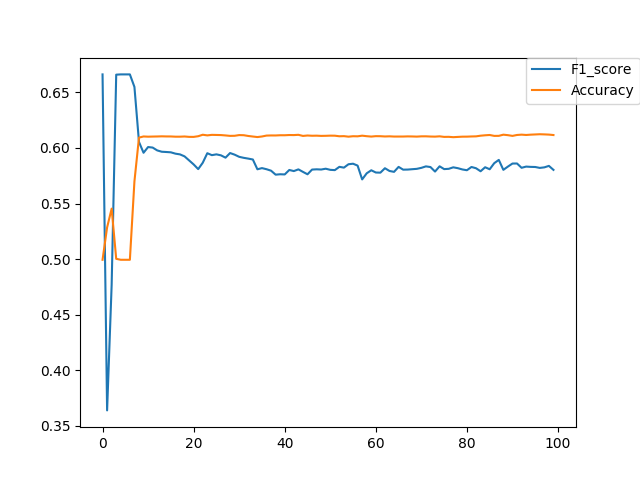}
\caption{F1 and Accuracy for Task 1}
\includegraphics[width=0.5\textwidth]{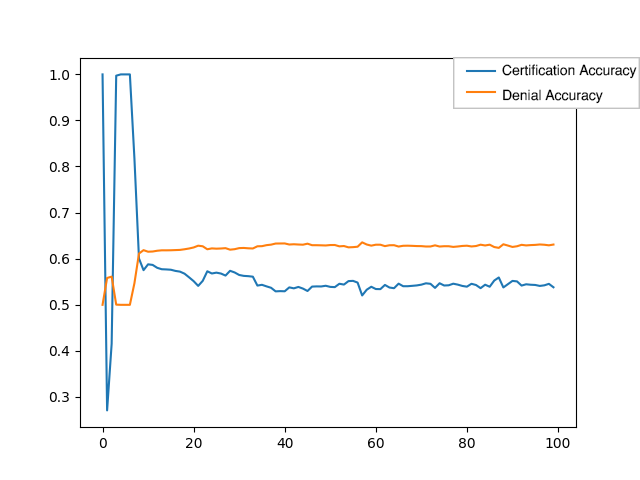}
\caption{Certification and Denial Accuracies for Task 1}
\end{figure}
The aim of this task is to predict the status of the case of an applicant, when all other attributes are provided. This task is a binary classification one for which we use the framework proposed in this paper. This data set causes a class imbalance problem (Class 1: 2.6 Million, Class 0: 180K). Hence, before applying the proposed model, the class imbalance problem is solved by randomly under-sampling the bigger class (In this case, Class 1). This reduced dataset is used for classification task using our model. The details of every stage are given below:
\begin{enumerate}
\item SOC\_NAME, FULL\_TIME\_POSITION, PREVAILING\_WAGE and YEAR attributes of the dataset have been divided into categorical classes after the pre-processing of the CSV. Thus by using the StringIndxer and OneHotEncoder of Spark, one hot vectors for each of these attributes is made for all the entries. These attributes are called CategoricalCols.
\item Soc\_prob and State\_prob both contain probabilities that are float values. These attributes are directly used for making the data frames. These attributes are called NumericCols.
\item Using the VectorAssembler of Spark, all the attributes in the CategoricalCols and NumericCols are combined into a single a vector and are stored in the same data frame with attribute name as Features.
\item The CASE\_STATUS classes are assigned a new attribute name Label which is used by the classifiers below.
\item For the new attributes, a Pipeline is created and the entire initial data frame is passed through the pipeline. 
\item The data frame is split into TrainingData and TestingData in the ratio 70:30.
\item Using the Features and Labels attributes, the initial classifier is trained. For this task, Logistic Regression is chosen for initial classification. After applying Logistic Regression, two new attributes are added:
\begin{enumerate}
\item \textit{Probability scores}, which contain the probabilities for both the classes that could be predicted by this classifier.
\item \textit{Prediction labels}, which contain the class predicted by this classifier.
\end{enumerate}
\item The probability score and prediction labels attributes are used as the knowledge gained from the dataset and are appended to the vectors in the Features.  
\item The new feature vectors are used further to train the multi-layer perceptron. The details of the perceptron used are as follows:
\begin{enumerate}
\item Each input layer has 1014 features, which is equivalent to the number of features in the new feature vectors. 
\item The perceptron has two hidden layers. The first hidden layer has 512 units and the second hidden layer has 128 units. For each of the units of hidden layers sigmoid activation function is used.
\item The output layers consists of two units, equivalent to the number of classes to be predicted. Softmax activation function is used for this layer
\item The perceptron is run for 100 epochs with a batch size of 1024.
\item The perceptron is trained using back-propagation with a logistic loss function.
\end{enumerate}
\item The prediction of the multi-layer perceptron is used as the final prediction of the model. The overall F1 score and accuracy are calculated.
\end{enumerate}
Results are shown in Fig. 2 and Fig. 3. Accuracy, F1 Score and Individual Class accuracies are shown in Table \ref{tab:modelperformances}. A summary of the artificial intelligence tools used in this task is presented in Table \ref{tab:modeldeets}.
\subsection{Task 2 - Classification on the basis of ``Prevailing\_Wage"}
The aim of this task is to predict whether the salary of an applicant would lie above the salary threshold set for the H-1B Visa or not, given all the other attributes. This task is a binary classification one which can be used by both the employer and the applicant. The current threshold for the two classes is set at \$90,000. All the applicants with a salary above the threshold are assigned Class 1 and the others are assigned Class 0. Only the applicants whose H-1B visa is certified are used in this task. This reduced the dataset into 2.6 Million entries. Most of steps are similar to Task 1, the changes made in the model for this task are mentioned below:

\begin{enumerate}
\item The new “Prevailing-Wage classes  are assigned according to the threshold value and is given  a new attribute name “Label” which is used by the classifiers below.
\item For this task, we chose Binomial Naive Bayes Classifier for initial classification. 
\item The multi-layer perceptron used for this task is a bit different from the one used in Task 1. The differences are as follows:

\begin{enumerate}
\item Each input layer has 1319 features, which is equivalent to the number of features in the new feature vectors. 
\item The perceptron has two hidden layers. The first hidden layer has 256 units and the second hidden layer has 32 units. For each of the units of hidden layers sigmoid activation function is used.
\end{enumerate}

\end{enumerate}
Results regarding F1 and accuracy of the model on this task are shown in Fig. 4.Accuracy, F1 Score and Individual Class accuracies are shown in Table \ref{tab:modelperformances}. A summary of the artificial intelligence tools used in this task is presented in Table \ref{tab:modeldeets}.
\vspace*{-1mm}
\begin{figure}
\hspace{-1mm}
\includegraphics[width=0.5\textwidth]{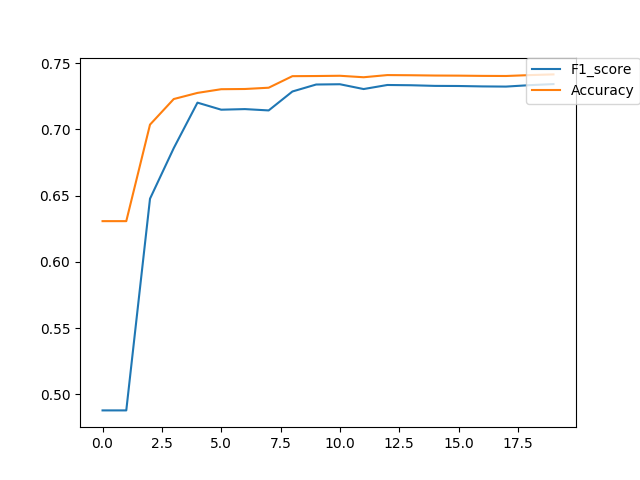}
\caption{F1 and Accuracy for Task 2}
\end{figure}

\begin{table*}[htb]
\begin{center}
\begin{tabular}{|c|c|c|c|c|}
\hline
 & \multicolumn{2}{|c|}{\textbf{AFTER STAGE 1}} & \multicolumn{2}{c|}{\textbf{AFTER STAGE 3}} \\
\hline
 & \textbf{F1 Score} & \textbf{Accuracy}& \textbf{F1 Score} & \textbf{Accuracy}\\
 \hline
\textbf{TASK 1 }& 0.5679 & 0.6006 & \textbf{0.5802} & \textbf{0.6116}\\
\textbf{TASK 2 }& 0.7261 & 0.7359 & \textbf{0.7352} & \textbf{0.7440}\\
\textbf{TASK 3 }& 0.4901 & 0.5078 & \textbf{0.5013}& \textbf{0.5201}\\
\textbf{TASK 4} & 0.6750 & 0.6030 & \textbf{0.6896} & \textbf{0.6315}\\
\hline
\end{tabular}
\end{center}
\caption{Performance metrics (F1 score and Accuracy) of the different task based models showing comparisons between the performance of only a Spark based framework against the proposed novel framework.}
\label{tab:modelperformances}
\end{table*}

\begin{table*}[htb]
\begin{center}
\begin{tabular}{|c|c|c|}
\hline
 & \textbf{Stage 1 Classifier} & \textbf{Stage 3 Deep Learning Layers}\\
 \hline
\textbf{TASK 1 }& Logistic Regression & [1013,528,128,2] \\
\hline
\textbf{TASK 2 }& Naive Bayes & [1319,256,32,2] \\
\hline
\textbf{TASK 3 }& Naive Bayes & [1319,256,32,4] \\
\hline
\textbf{TASK 4} & Logistic Regression & [281,64,64,2] \\
\hline
\end{tabular}
\end{center}
\caption{Analysis of the machine learning models used in stage 1, and the deep neural network used in stage 3.}
\label{tab:modeldeets}
\end{table*}

\subsection{Task 3 - Recommendation on the basis of “Prevailing-Wage”}
The aim of this task is to recommend the optimal salary range, within which the applicant should negotiate his/her joining income with the employer, given his/her application profile. This task is a multi-class recommender. The entire range of the Prevailing-wage is divided into four different ranges such that each range has equal number of entries. Only the certified applicants are considered in this model. Most of steps are similar to Task 2, the changes made in the model for this task are mentioned below:
\begin{enumerate}
\item The Label attribute of Task 2 is used only for Stage 1, from where the probabilities of the binary classifier is used for the Stage 2 of the model.
\item The new Prevailing-Wage classes are assigned according to the range in which the salary is and is given a new attribute name Label1 which is used by the multi-layer perceptron.
\item The multi-layer perceptron used is a bit different from the one used in Task 2. Four output units are used for the multi-layer perceptron so as to perform the multi-class classification. 
\end{enumerate}
Accuracy, F1 Score and Individual Class accuracies are shown in Table \ref{tab:modelperformances}. A summary of the artificial intelligence tools used in this task is presented in Table \ref{tab:modeldeets}.\\ \\



Now, we discuss the tasks run on dataset 2 or the \textbf{Arrhythmia Dataset}:

\subsection{Task 4: Classification of Cardiac Arrhythmia}
This dataset is taken from the UCI Machine Learning Repository and the problem is a binary classification problem which detects whether a patient has arrhythmia or not, given his/her attributes. The following is implemented for this task:

\begin{enumerate}
\item All 279 attributes are used as it is. All of these attributes are in the form of numerical values, hence all the attributes are used in the NumericCols.
\item The VectorAssembler is used to concatenate all the 279 attributes into a single vector. This vector attribute is placed under Features in the data frame.
\item The dataset contained 16 different types of arrhythmia, but as there are only 457 instances in the dataset, all arrhythmia instances are assigned Class 1 and all non-arrhythmia are assigned Class 0. These new class numbers are placed under Labels in the data frame.
\item Apart from the changes mentioned above, Stage 1 and Stage 2 of the model used is the same one used in Task 1 of the H-1B Visa Application dataset.
\item The multi-layer perceptron used for Stage 3 of the perceptron had the following attributes:
\item The input layer consists of 281 units.
\begin{enumerate}
\item Two hidden layers are used, with 64 units each \& Sigmoid activation used for each unit.
\item The output layer consists of 2 units, corresponding to the two class labels. Softmax activation is used for this layer.
\item The entire network is trained using back-propagation with a logistic loss function.
\item Training is done with a batch size of 50 and 100 epochs.
\end{enumerate}
\item The prediction of the multi-layer perceptron is used as the final prediction of the model. The overall F1 score and accuracy are calculated.
\end{enumerate}

\begin{figure}
\hspace{-1mm}
\includegraphics[width=0.5\textwidth]{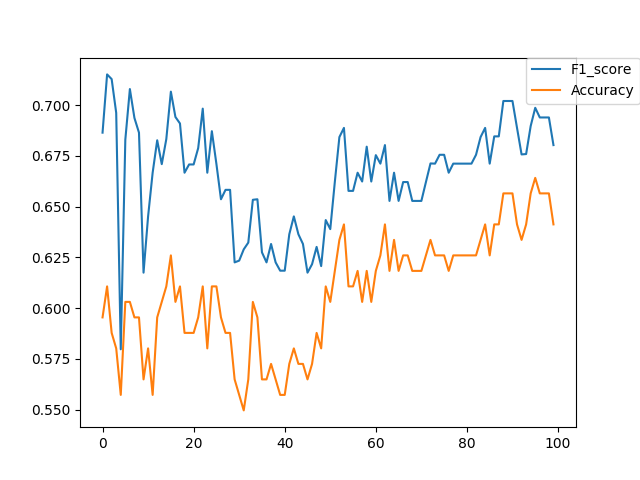}
\includegraphics[width=0.5\textwidth]{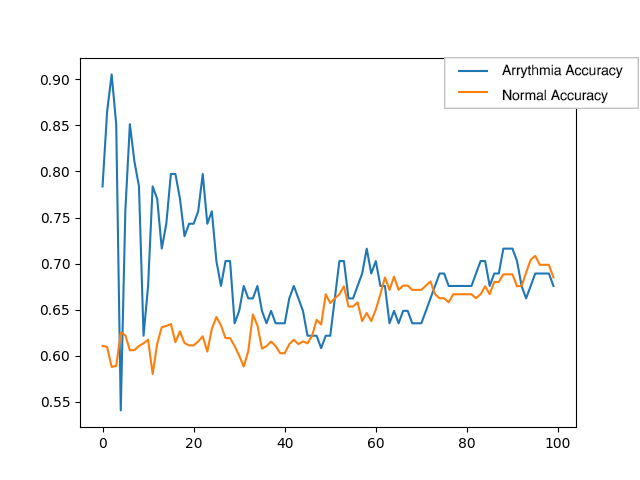}
\caption{Results for Task 4}
\end{figure}

Results are shown in Fig. 5. Accuracy, F1 Score and Individual Class accuracies are shown in Table \ref{tab:modelperformances}. A summary of the artificial intelligence tools used in this task is presented in Table \ref{tab:modeldeets}.

\section{CONCLUSIONS}

In this paper, we presented a novel framework for the analysis of big data. The proposed framework combined two widely popular tools, namely Apache Spark and Deep Learning, under the umbrella of a single structure. The connection between these tools was established using a third technique called Cascade Learning. This three-tier combination enabled us to conduct big data analysis with higher accuracy from a new perspective. By using these highly acclaimed individual tools in coherence with each other, we were able to obtain a model that is capable of conducting large scale big data analysis tasks within short periods of time, with lesser computational complexity and with significantly higher accuracy.  This model is an outer structure that allowed us to model all machine learning tasks, such as classification and recommendation, with ease. Our experiments on two real world datasets corroborated our claims of improved accuracy on varied machine learning setups, and hence enhanced the significance of the proposed methodology.

\addtolength{\textheight}{-12cm}

\section{SCOPE FOR FUTURE WORK}
While conducting the research for this paper, we faced some challenges. One key challenge was using the framework in a setup where both stages were multi-class. The initial stage, when run through a multi-class classifier and then combined with the multi-class deep learning stage yielded performances lower than its binary initial stage counterpart. We aim to overcome this challenge in future research.

\bibliography{make.bib}

\begin{thebibliography}{10}

\bibitem{zaharia2010spark}
M.~Zaharia, M.~Chowdhury, M.~J. Franklin, S.~Shenker, and I.~Stoica, ``Spark:
  Cluster computing with working sets.,'' {\em HotCloud}, vol.~10, no.~10-10,
  p.~95, 2010.

\bibitem{meng2016mllib}
X.~Meng, J.~Bradley, B.~Yavuz, E.~Sparks, S.~Venkataraman, D.~Liu, J.~Freeman,
  D.~Tsai, M.~Amde, S.~Owen, {\em et~al.}, ``Mllib: Machine learning in apache
  spark,'' {\em Journal of Machine Learning Research}, vol.~17, no.~34,
  pp.~1--7, 2016.

\bibitem{fu2016spark}
J.~Fu, J.~Sun, and K.~Wang, ``Spark--a big data processing platform for machine
  learning,'' in {\em Industrial Informatics-Computing Technology, Intelligent
  Technology, Industrial Information Integration (ICIICII), 2016 International
  Conference on}, pp.~48--51, IEEE, 2016.

\bibitem{nair2015streaming}
L.~R. Nair and S.~D. Shetty, ``Streaming twitter data analysis using spark for
  effective job search,'' {\em Journal of Theoretical and Applied Information
  Technology}, vol.~80, no.~2, p.~349, 2015.

\bibitem{nodarakis2016large}
N.~Nodarakis, S.~Sioutas, A.~K. Tsakalidis, and G.~Tzimas, ``Large scale
  sentiment analysis on twitter with spark.,'' in {\em EDBT/ICDT Workshops},
  pp.~1--8, 2016.

\bibitem{kotsiantis2006handling}
S.~Kotsiantis, D.~Kanellopoulos, P.~Pintelas, {\em et~al.}, ``Handling
  imbalanced datasets: A review,'' {\em GESTS International Transactions on
  Computer Science and Engineering}, vol.~30, no.~1, pp.~25--36, 2006.

\bibitem{sonak2016new}
A.~Sonak, R.~Patankar, and N.~Pise, ``A new approach for handling imbalanced
  dataset using ann and genetic algorithm,'' in {\em Communication and Signal
  Processing (ICCSP), 2016 International Conference on}, pp.~1987--1990, IEEE,
  2016.

\bibitem{guyon2003introduction}
I.~Guyon and A.~Elisseeff, ``An introduction to variable and feature
  selection,'' {\em Journal of machine learning research}, vol.~3, no.~Mar,
  pp.~1157--1182, 2003.

\bibitem{popescu2014feature}
M.~C. Popescu and L.~M. Sasu, ``Feature extraction, feature selection and
  machine learning for image classification: A case study,'' in {\em
  Optimization of Electrical and Electronic Equipment (OPTIM), 2014
  International Conference on}, pp.~968--973, IEEE, 2014.

\bibitem{dey2016paraphrase}
K.~Dey, R.~Shrivastava, and S.~Kaushik, ``A paraphrase and semantic similarity
  detection system for user generated short-text content on microblogs.,'' in
  {\em COLING}, pp.~2880--2890, 2016.

\bibitem{lavravc2010explicit}
N.~Lavra{\v{c}}, J.~F{\"u}rnkranz, and D.~Gamberger, ``Explicit feature
  construction and manipulation for covering rule learning algorithms,'' in
  {\em Advances in Machine Learning I}, pp.~121--146, Springer, 2010.

\bibitem{silva2008data}
L.~M. Silva, J.~M. de~S{\'a}, and L.~A. Alexandre, ``Data classification with
  multilayer perceptrons using a generalized error function,'' {\em Neural
  Networks}, vol.~21, no.~9, pp.~1302--1310, 2008.

\bibitem{sharma2014big}
C.~Sharma, ``Big data analytics using neural networks,'' 2014.

\bibitem{hu2010pattern}
Y.-C. Hu, ``Pattern classification by multi-layer perceptron using fuzzy
  integral-based activation function,'' {\em Applied Soft Computing}, vol.~10,
  no.~3, pp.~813--819, 2010.

\bibitem{pal1992multilayer}
S.~K. Pal and S.~Mitra, ``Multilayer perceptron, fuzzy sets, and
  classification,'' {\em IEEE Transactions on neural networks}, vol.~3, no.~5,
  pp.~683--697, 1992.

\bibitem{sheikh2015}
S.~M. Sarwar, M.~Hasan, and D.~I. Ignatov, ``Two-stage cascaded classifier for
  purchase prediction,'' {\em arXiv preprint arXiv:1508.03856}, 2015.

\bibitem{simon2016}
M.~Simonovsky and N.~Komodakis, ``Onionnet: Sharing features in cascaded deep
  classifiers,'' {\em arXiv preprint arXiv:1608.02728}, 2016.

\bibitem{christ2016automatic}
P.~F. Christ, M.~E.~A. Elshaer, F.~Ettlinger, S.~Tatavarty, M.~Bickel,
  P.~Bilic, M.~Rempfler, M.~Armbruster, F.~Hofmann, M.~D’Anastasi, {\em
  et~al.}, ``Automatic liver and lesion segmentation in ct using cascaded fully
  convolutional neural networks and 3d conditional random fields,'' in {\em
  International Conference on Medical Image Computing and Computer-Assisted
  Intervention}, pp.~415--423, Springer, 2016.

\bibitem{long2015fully}
J.~Long, E.~Shelhamer, and T.~Darrell, ``Fully convolutional networks for
  semantic segmentation,'' in {\em Proceedings of the IEEE Conference on
  Computer Vision and Pattern Recognition}, pp.~3431--3440, 2015.

\end{thebibliography}
\bibliographystyle{ieeetr}

\end{document}